\documentclass{article}
\usepackage{arxiv}
\usepackage[utf8]{inputenc} 
\usepackage[T1]{fontenc}    
\usepackage{hyperref}       
\usepackage{url}            
\usepackage{booktabs}       
\usepackage{amsfonts}       
\usepackage{nicefrac}       
\usepackage{microtype}      
\usepackage{lipsum}		
\usepackage{graphicx,grffile}

\usepackage{natbib}
\usepackage{doi}
\usepackage{algorithm}
\usepackage{algorithmic}
\newtheorem{theorem}{Theorem}

\newtheorem{obs}{Observation}
\newtheorem{corollary}{Corollary}

\title{Minimum Constraint Removal Problem for Line Segments is NP-hard}

\author{ \href{https://orcid.org/0000-0003-3387-3193}{
		\hspace{1mm}Bahram Sadeghi Bigham}
	\thanks{Corresponding Author: Bahram Sadeghi Bigham {https://iasbs.ac.ir/~b\_sadeghi\_b}}
	\\Department of Computer Science and Information Technology\\ Institute for Advanced Studies in Basic Sciences (IASBS)\\ Gava Zang, Zanjan, Iran. \\
	\texttt{b\_sadeghi\_b@iasbs.ac.ir} \\
}

\hypersetup{
	pdftitle={Minimum Constraint Removal Problem for Line Segments is NP-hard},
	pdfsubject={q-bio.NC, q-bio.QM},
	pdfauthor={Bahram Sadeghi Bigham},
	pdfkeywords={Algorithm, Computational Geometry, Robotics, Computer},
}

\begin{document}
	\maketitle
	\begin{abstract}
		In the minimum constraint removal ($MCR$), there is no feasible path to move from the starting point towards the goal and, the minimum constraints should be removed in order to find a collision-free path. It has been proved that $MCR$ problem is $NP-hard$ when constraints have arbitrary shapes or even they are in shape of convex polygons. However, it has a simple linear solution when constraints are lines and the problem is open for other cases yet. In this paper, using a reduction from Subset Sum problem, in three steps, we show that the problem is NP-hard for both weighted and unweighted line segments.
	\end{abstract}
	
	
	\section{Introduction}
		One of the most important objectives in motion planning is finding a feasible path from the starting point to a goal without collision with obstacles. The obstacles are either closed doors, which can be opened and removed by the robot, or are obstacles that cannot be passed which can be ignored by the robot with a penalty.\\
		Usually, there is no feasible path for some navigation. Recently, some researchers have focused on finding a path for the robot by minimizing the number of removed obstacles. For instance, in Stilman and Kuffner's paper \citep{stilman2005navigation}, the robot is able to move the obstacles around and clear its movement space. 
		In some studies, the main objective is to recognize the existence or absence of a feasible path
		\citep{basch2001disconnection}. The general problem is known as the Minimum Constraint Removal 
		($MCR$) and is introduced by Kris Houser in 2013 \citep{hauser2014minimum}.
		He proved that the discrete version of the $MCR$ problem with obstacles in arbitrary shapes is $NP-hard$. Later, in 2015 Houser in his paper titled "Minimum Constraint Displacement ($MCD$)" discussed a general problem of $MCD$ which its aim was changing the least number of possible obstacles to find a feasible path \citep{hauser2013minimum}.
		
		In this problem, if obstacles disappear without displacement, the $MCD$ problem will reduce to the $MCR$ problem. Lavalle and Ericsson in 2013, proved that when the obstacles are convex polygons, the $MCR$ problem is still $NP-hard$ \citep{erickson2013simple}.
		In 2015, Krontiris and Bekris used the approximation and searching algorithms to reduce the computational cost of $MCR$ \citep{krontiris2015computational}. In 2016, Huaqing Min and Bo Xu presented an approximation algorithm to solve the discrete-based $MCR$ problem \citep{xu2016solving}. They utilized a social-force-model based on the ant-colony technique and showed that their algorithm have a better performance in terms of the time and quality compared with the exact and greedy algorithms. 
		Also, Sergey Bereg and David Kirkpatrick in 2009 have presented two sub-problems, thickness and resilience, regarding wireless sensor network \citep{bereg2009approximating}. Covered regions of every sensor are considered as unit disks in the mentioned problem. By creating the dual graph from every given region, they showed that the thickness problem is equivalent to the length of the shortest path in dual graph. They further presented an approximation algorithm to solve it. Dual graph has many other applications which authors have used before \citep{bigham2012near} and also in this paper. Recently, in 2020, Sadeghi Bigham et al. \citep{bigham2020polynomial} solved the MCR problem for a special case in which all the constraints are axis-aligned-unit squares and the obstacles have only local effects. Local effect means there are no two cells which have the same label sets. They presented an $O(n^3)$ algorithm to solve it in the in the worst case.

		Clearly, the $MCR$ problem with only line-type obstacles can be solved in linear time by checking whether the start and goal points are at the same side of each line. The mentioned problems can be considered as hard and easy cases of the $MCR$ problem. However, there are other variant of the $MCR$ problem with obstacles with particular properties that have not been studied yet. In this paper, we discuss an $MCR$ problem in which the obstacles are line segments in a 2D closed environment. Specifically, we consider the problem with weighted and unweighted line segments. We prove that both type of the problem are NP-hard by reducing from the Subset Sum problem to them.
		
		The rest of this paper is organized as follows: Section 2 reviews previous work and different modeling for the $MCR$ problem. Section 3 discusses $MCR$ for weighted line segments and its extended version with unweighted line segments. The section also shows that the both problems are NP-hard and can be reduced from the Subset Sum problem. Section 4 describes conclusions and draws future work.
	
	\section{Minimum Constraint Removal Problem}
\begin{figure}
			\begin{center}
			\fbox{\includegraphics[scale=1.8]{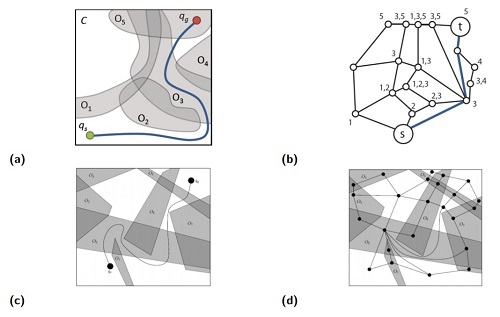}}
				\caption{\label{fourfig}$MCR$ problem with obstacles in arbitrary shapes (a), its graph model (b) \citep{hauser2014minimum}, $MCR$ for convex polygons (c) and related graph (d) \citep{hauser2013minimum}.}
			\end{center}
		\end{figure}

		The minimum constraint removal or briefness $MCR$ was first introduced by Houser in 2013 \citep{hauser2013minimum}. As it can be seen in Figure \ref{fourfig} (a and c), the gray regions represent obstacles and the rest of the figure represents feasible zones. $q_{s}$ and $q_{t}$ are the starting and target points, respectively. A point robot is in search of a path with minimum constraint from the start towards the goal point. In Figure \ref{fourfig}(a), the obstacles $O_{3}$ and $O_{5}$ must be removed from the path between $q_{s}$ and $q_{t}$ to create a feasible path.
		In every zone, one point is considered as the vertex of the graph. The vertices in the free zone are not labeled, while in the other zones, the label of overlapped obstacles are selected as the vertices labels. For instance, if a vertex is labeled with 5, it means it is covered by Obstacle 5. The starting and goal points in graph are shown with $s$ and $t$. Partitioning the continuous space creates the discrete $MCR$ on map and the $MCR$ problem aims to find the minimal subset of obstacles that covers all the vertices in the path between $s$ and $t$ (Figure \ref{fourfig}(b)).
		In the continuous case, the $d$-dimension configuration space
		$C$ is subset of $R^{d}$ (i.e., $C\subseteq{R^{d}}$), $n$ open sets of obstacles $S=\{s_{1},...,s_{n}\}$, the starting point $q _{s}$, and the target point $q_{t}\in C$ are given. The output is minimum constraint removal $S^{*}\subseteq S$. In discrete case, input is graph $G(V,E)$, cover function
		$c[v]$ and the starting and target points $s,t \in V$.
		For each vertex $v$, $c[v]$ has been annotated with a subset of
		$\{1,...,n\}$ which represents the number of obstacles available in a vertex $v$. The output is subset with minimum size $S^{*}$, so that a path $P$ in graph $G$ from $s$ to $t$ in terms of $c[v] \subseteq S^{*}$ for all $v\in P$ is available (refer to Figures \ref{fourfig}(b) and \ref{fourfig})(d).
		
		Lavalle and Erickson \citep{erickson2013simple} studied a special case of the $MCR$ problem and proved that the $MCR$ problem, is $NP-hard$ even for convex polygons (Fig. \ref{fourfig}(c)).
		Houser showed that the discrete version of the problem is $NP-hard$ by reduction from the $minimum$ $set-cover$ problem \citep{hauser2014minimum}. 
		\\
		For converting the $MCR$ problem with convex polygons obstacles into the form of discrete case, every connective zone which is intersection of a specific set of obstacles, specifies a vertex of the graph (Figure
		\ref{fourfig}(d)). The graph must be planar and the subgraphs created by the set of vertices which intersect with a special obstacle must be connected.
		The graph created by Houser which has been proved to be $NP-hard$ in discrete case, generally is not planar and the subgraphs created by sets of vertices which intersect with an obstacle are not connected. 
		
		In the next section, we discuss the $MCR$ problem when obstacles are line segments in a closed environment in $2D$.

	\section{MCR for Line Segments}
		In this section, we prove that $MCR$ problem for line segments is $NP-hard$. But, firstly $NP-hard$ness of two other problems will be showed. 
		
		It is clear that when all the obstacles are lines, then every line that $o$ and $g$ are at it's different side, should be removed and there is no need to remove the lines that $o$ and $g$ are at the same side of it. This statement is not valid when there exists line \textbf{segment}s in $S$. In this latter case, a line may be removed although $o$ and $g$ are at the same side of it (Observation \ref{Observation2}).In this case, and in the optimal path, it is possible that the robot crosses a line or line segment more than once and this makes the problem harder (Observation \ref{Observation1}).

\begin{figure}
			\centering
			\fbox{\includegraphics[scale=1]{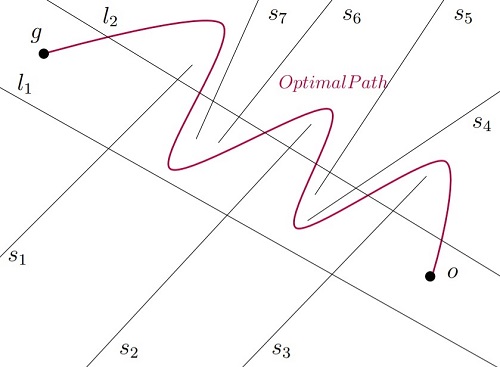}}
			\caption{\label{Observations}It is possible that the optimal path crosses a line more than once (Observation \ref{Observation1}). Also, a line may has to remove in the optimal path although both $o$ and $g$ are at the same side of it Observations \ref{Observation2}).}
		\end{figure}

		\begin{obs}
			The optimal path may be cross a line more than once (Fig. \ref{Observations}).
			\label{Observation1}
		\end{obs}
		
		\begin{obs}
			A line that both $o$ and $g$ are at its same side, should be removed in the optimal path (Fig. \ref{Observations}).
			\label{Observation2}
		\end{obs}
		
		Firstly, we define the decision version of the $MCR$ problem for weighted line segments in which each segment has different integer weight. In other words, removing each line segment has different cost. Problem 1 is called $MCR$ for Weighted Line Segments or $MCR-WLS$ and then a new problem (Problem 2) is introduced by adding some lines to the problem. This is an extended version of Problem 1 and is called $MCR-EWLS$. 
		
\begin{figure}
			\centering
				\fbox{\includegraphics[scale=1]{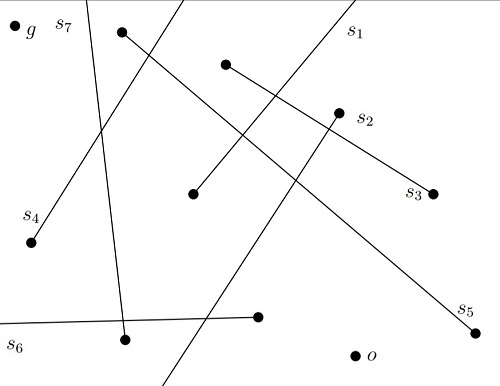}}
			\caption{\label{MCR-WLS}Robot wants to go from $o$ to $g$ and weighted line segments have blocked all the paths.}
		\end{figure}

		\textbf{Problem 1 (MCR-WLS)}: For $n$ given weighted line segments $S=\{s_1, s_2, ..., s_n\}$ in a rectangle with weight set $W=\{w_1, w_2, ..., w_n\}$, origin and goal points $o$ and $g$ and a constant $k$ (Fig. \ref{MCR-WLS}), find the path $\mathcal{P}$ from $o$ to $g$ which collides with segments with total weight $k$.

\begin{figure}
			\centering
		\fbox{\includegraphics[scale=1]{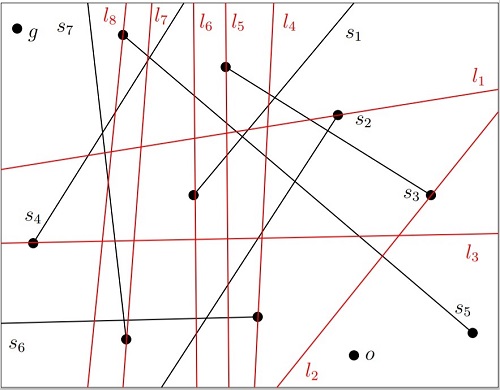}}
			\caption{\label{MCR-EWLS}7 line segments with new lines between $o$ and $g$ crossing endpoints.}
		\end{figure}

		\textbf{Problem 2 (MCR-EWLS)}: There are $n$ weighted line segments $S=\{s_1, s_2, ..., s_n\}$ with weight set $W=\{w_1, w_2, ..., w_n\}$, origin and goal points $o$ and $g$ and a constant $k$. Also, for every endpoint $i$ inside (not on the boundary) the rectangle (say $z$ endpoints) there is a line $l_i$ with weight $w(l_i)=M$ between $o$ and $g$ crossing the endpoint in which $M=1+\sum_{1}^{n} w(s_i)$ (Fig. \ref{MCR-EWLS}). The goal is finding a path from $o$ to $g$ with total weight $k+zM$.
		
		Let all the given segments have at least one endpoint inside the rectangle. We assume that for each segment $s_i$, there is (are) one (two) line(s) $l_{1i}$ (and $l_{2i}$) crossing its endpoint(s) and also the boundaries, in a way that $o$ and $g$ are at different sides of that (those) line(s).
		
		\begin{obs}\label{path}
			Every path $\mathcal{P}$ in $MCR-WLS$ with total weight $w(\mathcal{P})$ is equivalent to a path $\mathcal{P}'$ in $MCR-EWLS$ with weight $zM+w(\mathcal{P})$ in which $z$ is the the number of times that $\mathcal{P}$ intersects lines.
		\end{obs}
		
		\begin{figure}
			\centering
			\fbox{\includegraphics[scale=1]{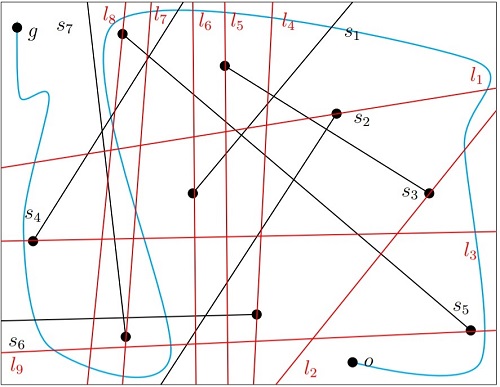}}
			\caption{\label{P:OPT}$P_{opt}$ from $o$ to $g$ crosses three line segments $s_1, s_4$ and $s_6$ with minimum weight 7.}
		\end{figure}
		
		In the following Theorem \ref{EWLS-NP-hard}, it is proved that Problem 2 is $NP-hard$ by a linear time reduction from known Subset Sum problem.  
		\begin{theorem}\label{EWLS-NP-hard}
			MCR-EWLS is $Np-hard$.
		\end{theorem}
		Proof: Let $A=\{a_1, a_2, ..., a_n\}$ includes $n$ integers and target sum $t$ are given. We want to find $A' \subseteq A$ which $\sum_{a_i \in A'}a_i=t$.

		\begin{figure}
			\centering
			\fbox{\includegraphics[scale=1]{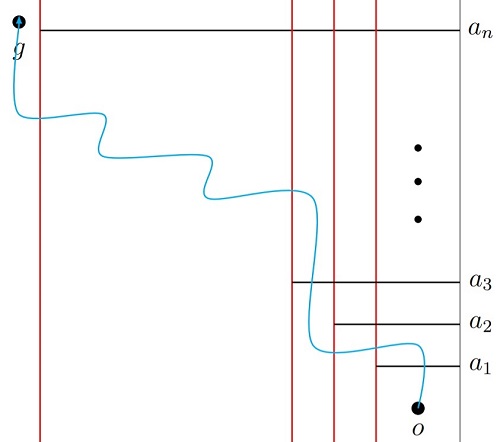}}
			\caption{\label{Reduction}Reduction from Subset Sum problem to MCR-EWLS problem.}
		\end{figure}
		
		In an axis aligned rectangle, we add a horizontal segment $s_i$ and a vertical line $l_i$ which first endpoint of segment is on the right edge of rectangle and the second endpoint is inside. Line $l_i$ passes throw the second endpoint of $s_i$ (Fig. \ref{Reduction}). 
		We put point $o$ to the most right-down cell and $g$ to the most left cell. If we set the weight of every segment $s_i$ to $w(s_i)=a_i$, then we have a problem in $MCR-EWLS$. In this problem, for all path $\mathcal{P}$ between $o$ and $g$, the weight of each vertical line is $M$ in which $M=1+\sum_{1}^{n} a(i)$. The goal is finding a path between $o$ and $g$ with total weight $nM+t$. 
		If there is a path with $w(\mathcal{P})=nM+t$, the the path crosses all the lines ($n$ lines) exactly once and it crosses some segments with total weight $t$ and so, it is a solution for given sample of Subset Sum problem. If the answer is no, and there is no solution for $MCR-EWLS$, then it is clear that there is no way to choose a subset of $A$ with total weight $t$.
		
		
		Now, by employing the Theorem \ref{EWLS-NP-hard}, one can conclude that Problem 1 ($MCR-WLS$) is $NP-hard$ as well.
		
		\begin{corollary}\label{WLS-NP-hard}
			$MCR-WLS$ problem is $NP-hard$.
		\end{corollary}
		
\begin{figure}
			\centering
			\fbox{\includegraphics[scale=1]{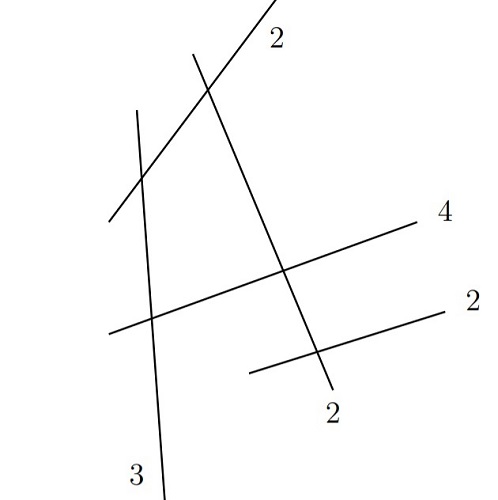}}
			\caption{\label{Unweighted1}Five weighted line segments in a 2D closed environment.}
		\end{figure}

		The final step includes a simple linear time reduction from $MCR-WLS$ in which each weighted line segment $s_i$ are copied $w_i$ times such that all copied segments have the same relation with the boundary and other obstacles. 
		\begin{theorem}\label{MCR-NP-hard}
			$MCR$ for line segments is $Np-hard$.
		\end{theorem}
		Proof: We need to show that $MCR-WLS \leq MCR-LS$, the problem for line segments, via a polynomial time reduction. Let $S=\{s_1, s_2, ..., s_n\}$ with integer weight set $W=\{w_1, w_2, ..., w_n\}$ and two origin and goal points $o$ and $g$ in a $2D$ closed rectangle are given. The goal is finding minimum total weight constraint $S'\subseteq S$ which a feasible path from $o$ to $g$ appears by removing $S'$ (Fig. \ref{Unweighted1}). 
		
\begin{figure}
			\centering
		\fbox{\includegraphics[scale=1]{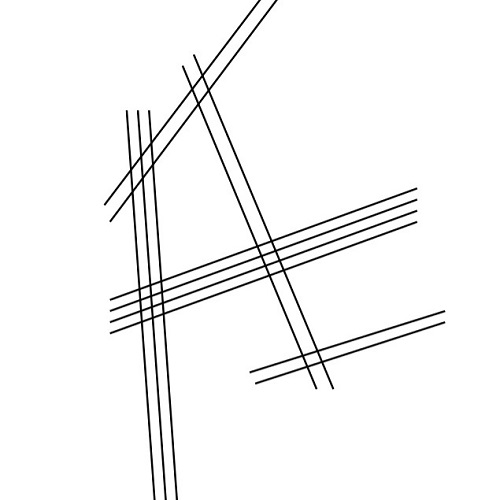}}
			\caption{\label{Unweighted2}Each line segment $i$ in Fig. \ref{Unweighted1} is copied $w_i$ times.}
		\end{figure}

		In this problem, for each line segment $s_i$, we duplicate it $w_i$ times which all copies have the same relation with the boundary and other line segments. In other words, they intersect other segments and boundary if $s_i$ intersects them (Fig. \ref{Unweighted2}). This reduction takes polynomial time even for segments with huge weights. Because if we consider $\cal{W}$ as the maximum weight of the segments, then we need at most $O(n\cal{W})$ duplication and the reduction is still polynomial. It is clear that if $w_i$ ($i\le n$) is not integer, then the problem is $NP-hard$ as well. 

	\section{Conclusion and Future Work}
		Minimum constraint removal ($MCR$) problem when all obstacles are lines in a plane, have a linear algorithm and when obstacles are arbitrary shapes or even convex polygons, is $NP-hard$. This problem was open for line segments, discs, fat shapes and some other types of obstacles. In this paper, we proved $NP-hard$ness of the problem for line segments in three steps. First, we defined a new problem ($MCR-EWLS$) and proved it is NP-hard by reduction from the known Subset Sum problem. Then, we showed that this problem is equivalent with the $MCR-WLS$ problem and then we concluded that the MCR problem for weighted line segments is $NP-hard$. At the final step, we presented a reduction from the $MCR-WLS$ to our main problem and showed that $MCR$ problem for line segments is $NP-hard$.
		
		For future work, we are interested in proving that $MCR$ problem in $2D$ is $NP-hard$ even for fat shapes. Also, one can present approximation or heuristic algorithms to solve new types of NP-hard cases of the problem.

	\bibliographystyle{unsrtnat}
	\bibliography{template} 
\end{document}